\documentclass [12pt]{iopart}                                                                                               
\usepackage{iopams}                                                           
\begin{document}
\newtheorem{Theorem}{Theorem}
\newtheorem{Proposition}{Proposition}
\newtheorem{Lemma}{Lemma}                                                           
\newtheorem{Definition}{Definition}  
\newtheorem{Corollary}{Corollary} 
\newtheorem{Remark}{Remark}  
\eqnobysec


\title{Spherically symmetric steady states of galactic dynamics in scalar gravity}
\author{Simone Calogero}
\address{
Institut f\"ur Mathematik der Universit\"at Wien,\\
Strudlhofgasse 4, 1090 Vienna, Austria\\
E-mail: caloges3@univie.ac.at}

\begin{abstract}
The kinetic motion of the stars of a galaxy is considered within the framework of a relativistic scalar theory of gravitation. This model, even
though unphysical, may represent a good \textit{laboratory} where to study in a rigorous, mathematical way those problems,
like the influence of the gravitational radiation on the dynamics, which are still
beyond our present understanding of the physical model represented by the Einstein--Vlasov system. The present paper is devoted to derive the
equations of the model and to prove the existence of spherically symmetric equilibria with finite radius. 
\end{abstract}

\section{Introduction}
The non-relativistic kinetic model for a collisionless ensemble of particles 
interacting through the gravitational forces that they generate collectively
is governed by the Vlasov--Poisson system:
\begin{eqnarray}
\partial_t f+v\cdot\nabla_xf-\nabla_xU\cdot\nabla_vf=0,\\
\Delta U=4\pi\rho,\\
\rho (t,x)=\int\rmd v f(t,x,v).
\end{eqnarray}
The Vlasov--Poisson system can be used to describe the motion of the stars in a galaxy, thought of as point particles, when the relativistic
effects are negligible.
In this application, the function $f(t,x,v)$ in the previous equations is non-negative and gives the density on phase space of
the stars in the galaxy, where $t\in\mathbb{R}$ denotes the time and $x,v\in\mathbb{R}^3$ are the position and the velocity respectively. The
function $U=U(t,x)$ is the mean Newtonian potential generated by the stars.

When the relativistic effects are not negligible, the motion of the stars has to be described by the Einstein--Vlasov system. In this case, the
dynamics of the matter is still described by a non-negative scalar function $f$, but the unknown of the field equations is now a symmetric
tensor of second rank, the metric of the space-time.

The Einstein--Vlasov system is much more complicated than Vlasov--Poisson. The greatest difficulties arise from the character of the Einstein
equations, which are essentially hyperbolic and highly non-linear even in the absence of sources, from the equivalence of all the coordinate
systems in general relativity and from the fact that in the Einstein theory of gravitation, the space-time is not given in advance but is
itself part of the solution of the equations. All these features make it extremely difficult to obtain global solutions of the Einstein
equations coupled to any kind of matter. For instance, while global solutions of Vlasov--Poisson are known to be launched by general data
of the Cauchy problem (see \cite{Pf}), in the case of Einstein--Vlasov this result is known only for spherically symmetric solutions with 
small initial data (see \cite{RR}). In particular, the assumption of spherical symmetry rules out one of the most interesting new features of 
general relativity, which is the propagation of gravitational waves,
since this restriction forces the space-time to be static outside the support of the matter. It is thus not surprising that a 
rigorous mathematical theory of the gravitational radiation in general relativity is at present still missing, despite of the fact that 
its detection is currently one of the most challenging goals of experimental physics.   

In this paper we propose a simple, unphysical model in which the dynamics of the matter is still described by the Vlasov equation
but where the gravitational forces 
between the particles are now supposed to be mediated by a scalar field. In spite of the very simplified assumptions, 
this model is not by any means trivial, even when it is further reduced to the spherically symmetric case. 
In fact a scalar gravitation theory allows also for the propagation of spherical gravitational waves.
(This fact turns out to be very useful also for numerical investigations of the radiation in a scalar gravitation theory, see \cite{ST}).
For this reason, this model may represent a comparatively easier framework where to study the effects of the gravitational radiation on the
dynamics of a many particles system.
It is important to underline however that the results obtained for this scalar model might not be representative in general for the physical model
represented by the  Einstein--Vlasov system.    

We will refer to this scalar model as the Nordstr\"om--Vlasov system, since the scalar gravitation theory which we use to describe 
the interaction among the particles reduces, in a particular frame, to the one introduced by Nordstr\"om in \cite{No} (see remark 1 below).  
The equations will be derived in the next section by appealing to the more general case 
considered in \cite{DEF}. 
In section 3 we give a few preliminary results on the Nordstr\"om--Vlasov system for which the proof traces back to the case of 
the Vlasov--Poisson system. All these results concern the spherically symmetric stationary solutions and are obtained by adapting the ideas 
of \cite{BFH,RR2}.

For notational convenience, we study the model only for a single species of particle with unit mass (the generalization to the case 
of a mixture is straightforward). We also set the gravitational constant and the speed of light equal to one and remove the constant $4\pi$ 
from the equations. 
The convention for the signature of the metric is $- + + +$. This affects the definition in Cartesian coordinates of the wave operator in
Minkowski space, which becomes
\begin{equation}
\square=-\partial_t^2+\Delta.
\end{equation}


\section{The Nordstr\"om--Vlasov system}

The model which we are going to present describes the collective motion of collisionless particles interacting by means of their own self-generated gravitational
forces, under the condition that the dynamics of the gravitational field is described in accordance to a simple scalar gravitation metric
theory.  

By \textit{scalar} gravitation \textit{metric} theory we mean a theory in which the gravitational forces are 
mediated by a scalar field $\phi$ and the effect of such forces is to induce a curvature in the space-time. Moreover it is assumed that
the scalar field modifies the otherwise flat metric only by a rescaling. Therefore the physical metric in this theory will be
conformally flat, that is given by
\begin{equation}
g_{\mu\nu}=A^{2}(\phi)\eta_{\mu\nu},
\end{equation}
where $A$ is a strictly positive function and, adopting Cartesian coordinates, $\eta_{\mu\nu}=\textnormal{diag}(-1,1,1,1)$ (the use of a
Cartesian frame is clearly a natural choice, due to the presence of a background flat metric).
The metric $g_{\mu\nu}$ is assumed to be the physical one in the sense that it is the metric to which the matter will be coupled. 
The condition that the particles make up a collisionless ensemble in the space-time is carried out by requiring that the particle density 
be a solution of the Vlasov equation 
\begin{equation}
\partial_{t}f+\frac{p^{a}}{p^{0}}\partial_{x^{a}}f-\Gamma_{\mu\nu}^{a}\frac{p^{\mu}p^{\nu}}{p^{0}}\partial_{p^{a}}f=0,
\end{equation}
where $\Gamma_{\mu\nu}^{\sigma}$ are the Christoffel symbols of the physical metric, $f$ is the particle density and $p^{0}$ is 
determined by $p^{a}$ ($a=1,2,3)$ according to the relation $g_{\mu\nu}p^{\mu}p^{\nu}=-1$, which expresses the condition
that the four momentum $p^{\mu}$ lies on the mass shell of the physical metric (the Greek indexes run from $0$ to $3$). 
The previous equation is equivalent to postulate that the
function $f$ is constant on the geodesic flow of the metric (2.1). 
The symbols $\Gamma_{\mu\nu}^{\sigma}$ for the metric (2.1) are given by
\begin{displaymath}
\Gamma_{\mu\nu}^{\sigma}=\alpha(\phi)(\delta_{\nu}^{\sigma}\partial_{\mu}\phi+\delta_{\mu}^{\sigma}\partial_{\nu}\phi-
\eta_{\mu\nu}\partial^{\sigma}\phi),
\end{displaymath}
where $\alpha(\phi)=A^{-1}\frac{\rmd A}{\rmd\phi}$.
Hence, the Vlasov equation in a conformally flat space-time takes the form
\begin{equation}
\partial_{t}f+\frac{p^{a}}{p^{0}}\partial_{x^{a}}f-\alpha(\phi)[2(p^{\mu}\partial_{\mu}\phi)\frac{p^{a}}{p^{0}}+\frac{A^{-2}(\phi)}{p^{0}}
\partial^{a}\phi]\partial_{p^{a}}f=0.
\end{equation}

We remember at this point that two tensorial quantities play an important role for the Vlasov equation in a curved space-time. 
They are the current density,
\begin{equation}
N^{\mu}=-\int \frac{\rmd p}{p_{0}}\sqrt{\mathfrak{g}}\,p^{\mu}f
\end{equation}
and the stress-energy tensor,
\begin{equation}
T^{\mu\nu}=-\int \frac{\rmd p}{p_{0}}\sqrt{\mathfrak{g}}\,p^{\mu}p^{\nu}f.
\end{equation}
$\mathfrak{g}$ denotes the determinant of the metric and therefore, for the metric (2.1), we have
$\sqrt{\mathfrak{g}}=A^{4}(\phi)$.
The importance of $N^{\mu}$ and $T^{\mu\nu}$ is due to the fact that they are both divergence free:
\begin{eqnarray}
\nabla_{\mu}T^{\mu\nu}=0,\\
\nabla_{\mu}N^{\mu}=0,
\end{eqnarray}
where $\nabla_{\mu}$ is the covariant derivative of the metric $g_{\mu\nu}$. The relations (2.6)-(2.7) hold independently from the 
dynamics of the physical metric. They are consequences of (2.2) alone (see \cite{E}).

In order to define completely the dynamics, we need now to specify the coupling function $A(\phi)$ and to postulate an equation for 
${\phi}$.
To this purpose, we remember that in a second order theory where the gravitational effects are mediated both by a tensor field
and a scalar field, the field equations can be put in the form (see \cite{DEF},
equations (2.8)-(2.9)):
\begin{eqnarray}
E_{*}^{\mu\nu}=2T_{*}^{\mu\nu}+2\gamma(\phi)(\partial^{\mu}\phi\partial^{\nu}\phi+\frac{1}{2}g_{*}^{\mu\nu}g^{\rho\sigma}_{*}
\partial_{\rho}\phi\partial_{\sigma}\phi) - 2B(\phi)g_{*}^{\mu\nu},\\
\square_{g^{*}}\phi+\gamma(\phi)g_{*}^{\mu\nu}\partial_{\mu}\phi\partial_{\nu}\phi=-\alpha(\phi)T_{*}+\frac{dB(\phi)}{d\phi},
\end{eqnarray} 
where $E^{*}_{\mu\nu}$ is the Einstein tensor, $B$ the self coupling potential of the field $\phi$, $\gamma$ an arbitrary function of
$\phi$ and $\square_{g^{*}}$ the wave operator with respect to the metric $g^{*}_{\mu\nu}$ .
All the quantities in the previous equations with the suffix $*$ are expressed in terms of the metric $g^{*}_{\mu\nu}$, also known as the 
Einstein metric. However, the Einstein frame is not the physical one. The physical metric, or Fierz metric, is related to $g^{*}_{\mu\nu}$ 
by the conformal transformation  ${g}_{\mu\nu}= A^{2}(\phi)g^{*}_{\mu\nu}$, whereas the stress-energy tensor in the Fierz frame is given by
$T^{\mu\nu}=A^{-6}(\phi)T_{*}^{\mu\nu}$. The former rescaling law follows by the definition of the stress energy tensor 
(independent of the frame) as the variation with respect to the metric of
the matter action (see equation (2.16) of \cite{DEF}).
The advantage of using the Einstein frame instead of the Fierz frame is due to the fact that the 
dynamics of the metric and of the field are separate from each  other in the former, whereas in the latter they  turn out to be entangled. 

Equations (2.8)-(2.9) actually define a general class of Tensor Scalar theories, since they contain
the arbitrary function $\gamma(\phi)$. Among these theories, a special one is singled out by requiring the property of 
\textit{scale invariance}. 
This
means that there exists in the theory a continuous one parameter symmetry group whose action consists in rescaling the coupling function
$A(\phi)$ by a constant factor. This assumption determines completely 
the arbitrary functions of the theory (except a constant which we fix equal to one) according to the relations
\begin{displaymath}
A(\phi)=\rme^{\phi},\quad B(\phi)=\gamma(\phi)=0
\end{displaymath}  
(see \cite{DEF}, equations (2.19a-b-c)).
In this special theory the equation for the scalar field $\phi$ takes the form
\begin{equation}
\square_{g^{*}}\phi=-T_{*}.
\end{equation}

Let us explain now what we mean by \textit{simple} scalar gravitation theory: we assume that the scalar theory satisfies the scale 
invariance property. Motivated by the more general case indicated above, we then replace $A(\phi)$ in equation (2.1) with $\rme^{\phi}$. 
Therefore the Fierz metric in this scalar theory is given by
\begin{equation}
g_{\mu\nu}=\rme^{2\phi}\eta_{\mu\nu}
\end{equation}
and $\sqrt{\mathfrak{g}}=\rme^{4\phi}$. Accordingly, equation (2.3) becomes
\begin{equation}
\partial_{t}f+\frac{p^{a}}{p^{0}}\partial_{x^{a}}f-[2(p^{\mu}\partial_{\mu}\phi)\frac{p^{a}}{p^{0}}+\frac{\rme^{-2\phi}}{p^{0}}
\partial^{a}\phi]\partial_{p^{a}}f=0
\end{equation}
and the mass shell condition reads
\begin{equation}
p^{0}=\sqrt{\rme^{-2\phi}+\delta_{ab}p^{a}p^{b}}.
\end{equation}
We now introduce, beside the tensor (2.5), the ``unphysical'' stress energy tensor
\begin{equation}
T_{*}^{\mu\nu}=\rme^{6\phi}T^{\mu\nu} \Rightarrow T_{*}=\rme^{4\phi}T
\end{equation}
and postulate for $\phi$ the equation
\begin{equation}
\square\phi=- T_{*},
\end{equation}
where $\square$ is the wave operator in the flat space-time, given by (1.4). 
\begin{Remark}\textnormal{
In \cite{No}, the Finnish physicist Gunnar Nordstr\"om proposed a relativistic scalar theory of gravitation which was based on the field equation
(2.15). For this reason from now on we will refer to our model as the Nordstr\"om--Vlasov system
(or NV system for short).} 
\end{Remark}

In the Fierz frame and for a stress-energy tensor given by (2.5), equation (2.15) reads explicitly
\begin{equation}
\square\phi=\rme^{6\phi}\int \frac{\rmd p}{p^{0}}f(t,x,p).
\end{equation}

The Nordstr\"om--Vlasov system corresponds to the set of equations (2.12), (2.13), (2.16). Even though this system describes perfectly 
the model that we have in mind, it is convenient to rewrite it completely in the Einstein frame. We notice indeed that the 
particle density is still defined on the mass shell of $g_{\mu\nu}$. In order to remove even this last connection with the 
Fierz frame, we rescale the momentum as $p^{\mu}_{*}=\rme^{\phi}p^{\mu}$ and define the ``unphysical'' particle density by
\begin{equation}
f_{*}(t,x,p_{*})=f(t,x,\rme^{-\phi}p_{*}).
\end{equation}
The Nordstr\"om--Vlasov system in the Einstein frame then becomes
\begin{eqnarray}
\square\phi=\rme^{4\phi}\int \frac{\rmd p_{*}}{p^{0}_{*}}f_{*}(t,x,p_{*}),\\
p^{0}_{*}=\sqrt{1+\delta_{ab}p^{a}_{*}p^{b}_{*}},\qquad\qquad\qquad\qquad\qquad\qquad\qquad \textrm{(NV$_{*}$)}\\
\partial_{t}f_{*}+\frac{p_{*}^{a}}{p_{*}^{0}}\partial_{x^{a}}f_{*}-\frac{1}{p_{*}^{0}}(p_{*}^{\mu}\partial_{\mu}\phi
p_{*}^{a}+\partial^{a}\phi)\partial_{p_{*}^{a}}f_{*}=0.
\end{eqnarray}

To conclude this section, we derive from (2.6) and (2.7) the conservation of energy and of the total number of particles.
Unlike Einstein's general relativity theory, in fact, the scalar gravitation theory under discussion here offers a natural definition for the local
energy, in virtue of the presence of the background metric $\eta_{\mu\nu}$. 
By expanding equation (2.7) we get
\begin{displaymath}
\rme^{-4\phi}\partial_{\mu}\rme^{4\phi}N^{\mu}=0.
\end{displaymath}
The previous relation shows that the quantity $\partial_{t}(\rme^{4\phi}N^{0})$ is a pure divergence and therefore
\begin{displaymath}
\partial_{t}\int \rmd xN^{0}\rme^{4\phi}=0.
\end{displaymath}
Using (2.4) we then have
\begin{equation}
N=\int \rmd x\int \rmd p\, \rme^{6\phi}f(t,x,p)=\textrm{constant}.
\end{equation}
The identity (2.21) expresses the conservation of the total number of particles. The same relation in the NV$_*$ formulation of the model 
takes the form 
\begin{equation}
N_{*}=\int \rmd x\int \rmd p_{*}\rme^{3\phi}f_{*}(t,x,p_{*})=\textrm{constant}.
\end{equation} 
Likewise expanding (2.6) we get 
\begin{equation}
\partial_{\mu}T^{\mu\nu}+6\partial_{\mu}\phi T^{\mu\nu}-\rme^{-2\phi}\partial^{\nu}\phi T=0.
\end{equation}
By means of (2.14) and (2.15),
\begin{displaymath}
T=\rme^{-4\phi}T_{*}=-\rme^{-4\phi}\square\phi.
\end{displaymath}
Substituting into (2.23) and using the identity
\begin{displaymath}
\partial^{\nu}\phi\square\phi=
\partial_{\mu}(\partial^{\mu}\phi\partial^{\nu}\phi-\frac{1}{2}\eta^{\mu\nu}\partial^{\rho}\phi\partial_{\rho}\phi),
\end{displaymath}
we get
\begin{equation}
\rme^{-6\phi}\partial_{\mu}(\rme^{6\phi}T^{\mu\nu}+\partial^{\mu}\phi\partial^{\nu}\phi-\frac{1}{2}\eta^{\mu\nu}\partial^{\rho}\phi
\partial_{\rho}\phi)=0.
\end{equation}
For $\nu=0$, the previous equation leads to the conservation law
\begin{displaymath}
\partial_{t}\int \rmd x[\rme^{6\phi}T^{00}+\frac{1}{2}(\partial_{t}\phi)^{2}+\frac{1}{2}(\nabla\phi)^{2}]=0,
\end{displaymath}
which, using (2.5), implies
\begin{equation}
\mathcal{E}(t)=\int \rmd x\int \rmd p \, \rme^{8\phi}p^{0}f+\frac{1}{2}\int \rmd x[(\partial_{t}\phi)^{2}+(\nabla\phi)^{2}]=\textrm{constant}.
\end{equation}
For the NV$_*$ system the conservation of energy reads 
\begin{equation}
\mathcal{E}_{*}(t)=\int \rmd x\int \rmd p_{*}\rme^{4\phi}p_{*}^{0}f_{*}+
\frac{1}{2}\int \rmd x[(\partial_{t}\phi)^{2}+(\nabla\phi)^{2}]=\textrm{constant}.
\end{equation}

\section{Stationary spherically symmetric models}
We begin our investigation of the NV model by looking at the stationary spherically symmetric solutions. For this purpose we will work
with the NV$_{*}$ formulation. 
For notational convenience we rewrite below the equations removing
the suffix ${*}$ (in order to not generate confusion we continue to call this system NV$_{*}$).
\begin{eqnarray}
-\partial_t^2\phi+\Delta\phi=\rme^{4\phi}\int\frac{\rmd p}{\sqrt{1+p^{2}}}f(t,x,p),\\
\partial_{t}f+\frac{p}{\sqrt{1+p^{2}}}\cdot\partial_{x}f-[(\partial_{t}\phi+\frac{p\cdot\nabla\phi}{\sqrt{1+p^{2}}})p+
\frac{\nabla\phi}{\sqrt{1+p^{2}}}]\cdot\partial_{p}f=0.
\end{eqnarray}
(In this section we denote by $p$ the vector $p=(p_1,p_2,p_3)$ and put $p^2=|p|^2$).
 
Spherical symmetry means that 
$\phi=\phi(t,r)$ and $f(t,x,p)=f(t,Ax,Ap)$ for every orthogonal $3\times 3$ matrix $A$ and $x,p\in\mathbb{R}^{3}$. 
With this restriction, NV$_{*}$ becomes
\begin{eqnarray}
-\ddot{\phi}+\frac{1}{r^{2}}\frac{\rmd}{\rmd r}(\phi'r^{2})=\rme^{4\phi}\mu(t,r),\\
\partial_{t}f+\frac{p}{\sqrt{1+p^{2}}}\cdot\partial_{x}f-[(\dot{\phi}+\frac{\phi'}{r}\frac{x\cdot
p}{\sqrt{1+p^{2}}})p+\frac{\phi'}{r}\frac{x}{\sqrt{1+p^{2}}}]\cdot\partial_{p}f=0,\\
\mu(t,r)=\int \frac{\rmd p}{\sqrt{1+p^{2}}}f(t,x,p),
\end{eqnarray}
where ``dots'' and ``primes'' denote the temporal and the radial derivatives respectively.
We denote by NV$_*[r]$ the system of equations which defines the stationary (i.e. independent of $t$) solutions of (3.3)--(3.5):
\begin{eqnarray}
\frac{1}{r^{2}}\frac{\rmd}{\rmd r}(\phi'r^{2})=\rme^{4\phi}\mu(r),\quad\quad\\
p\cdot\partial_{x}f-\frac{\phi'}{r}[(x\cdot p)p+x]\cdot\partial_{p}f=0,\qquad\qquad\qquad\textnormal{(NV$_*[r]$)}\\
\mu(r)=\int \frac{\rmd p}{\sqrt{1+p^{2}}}f(x,p).
\end{eqnarray}

For later convenience we recall here the definition \textit{in the Einstein frame} of some important quantities:
\begin{eqnarray}
\rho(r)=\rme^{4\phi(r)}\int \rmd p \sqrt{1+p^{2}}f(x,p)\quad \textnormal{(mass-energy density)},\\
\mathcal{P}(r)=\rme^{4\phi(r)}\int \frac{\rmd p}{\sqrt{1+p^{2}}}\Big (\frac{x\cdot p}{r}\Big)^{2}f(x,p)\quad\textnormal{(radial pressure)},\\
\mathcal{P}_{T}(r)=\frac{1}{2}\rme^{4\phi(r)}\int \frac{\rmd p}{\sqrt{1+p^{2}}}\Big |\frac{x\wedge p}{r}\Big |^{2}f(x,p)\quad 
\textnormal{(tangential pressure)}
\end{eqnarray}
and note that (3.6) can be rewritten as
\begin{equation}
\frac{1}{r^{2}}\frac{\rmd}{\rmd r}(\phi'r^{2})=\rho(r)-(\mathcal{P}(r)+2\mathcal{P}_{T}(r)).
\end{equation}
We split our analysis in three different subsections. 
\subsection{The Jeans theorem}
In any stationary solution of a stellar dynamics model, whether it is the Vlasov--Poisson, the Einstein--Vlasov or the relativistic scalar
model under
discussion here, the particle density is a functional of the invariants of the associated Vlasov equation. The statement according to
which in the spherically symmetric case any of these invariants is in its turn a functional of the local energy and the angular momentum,
so that in the end the particle density depends upon these two integrals of the characteristic system only, is known as the ``Jeans
theorem''. It was established for the Vlasov--Poisson system (see \cite{BFH}) while for the Vlasov--Einstein model it was proved to be false 
(cf. \cite{S}). (It remains open the possibility that a modified formulation of the theorem could lead to a true statement). In the case 
of the  NV$_{*}$ system the Jeans theorem is easily seen to be valid. For we introduce the new coordinates
\begin{displaymath}
r=|x|,\quad w=\rme^{\phi}\frac{x\cdot p}{|x|},\quad F=\rme^{2\phi}(|x|^{2}|p|^{2}-(x\cdot p)^{2}),
\end{displaymath}    
which have value in the set
$G=\{(r,w,F): r>0, w\in\mathbb{R}, F>0\}$. (Strictly speaking, the variable $F$ could take also value zero. However, since $F$ is an integral of
motion and the invariant manifold defined by $F=0$ has zero measure, we can restrict ourselves to study the motion on $G$).
Since $f$ is spherically symmetric, it can be written in function of $r,w,F$, i.e.
\begin{displaymath}
f=f(r,w,F),\quad r>0, F>0, w\in\mathbb{R}.
\end{displaymath}
The equation (3.7) in these new coordinates takes the form
\begin{equation}
w\partial_{r}f+(\frac{F}{r^{3}}-\rme^{2\phi}\phi')\partial_{w}f=0.
\end{equation}
The characteristic system for the previous equation has the form considered in \cite{BFH} (see (2.11), pag. 162) with $m(r)=\rme^{2\phi}(\phi'
r^{2})$. Provided that $\phi$ is a continuously differentiable solution of (3.6), the function $m(r)$ is continuous and monotonically
increasing for $r>0$. Thus we can apply the result of \cite{BFH} and conclude that any measurable function which is constant on the 
characteristics of (3.7) has the form $f(r,w,F)=\Phi(\tilde{E},F)$, where 
\begin{displaymath}
\tilde{E}(r,w,F)=\frac{1}{2}w^{2}+\frac{1}{2}\frac{F}{r^{2}}+U(r)=
\frac{1}{2}\rme^{2\phi}p^{2}+U(r)
\end{displaymath}
and $U(r)$ is any primitive of the function $g(r)=\rme^{2\phi}\phi'$, say $U(r)=\frac{1}{2}\rme^{2\phi}$. 
With regard to NV$_*[r]$, this leads immediately to the following
\begin{Theorem}
Let $\phi\in C^1$ solves (3.6) (in a weak sense) and $f$ be measurable and constant on the characteristics of (3.7). 
Then there exists a unique $\Phi:\mathbb{R}^{2}\to [0,+\infty)$ 
such that 
\begin{equation}
f(x,p)=\Phi(E,F),
\end{equation}
where $E$ is the local energy of the particles, i.e. $E=\rme^{\phi}\sqrt{1+p^{2}}$, and $F$ the modulus squared of the 
local angular momentum, $F=\rme^{2\phi}|x\wedge p|^{2}$.
\end{Theorem}

Clearly, if $\Phi$ is $C^{1}$, then $f$ solves (3.7) classically.

Substituting (3.14) into the definitions (3.9)-(3.11) we get, after a transformation of variables,
\begin{eqnarray}
\rho(r)=\frac{\pi}{r^{2}}\int_{\rme^{\phi}}^{\infty}\rmd E\,E^{2}\int_{0}^{r^{2}(E^{2}-\rme^{2\phi})}\rmd F\frac{\Phi(E,F)}
{\sqrt{E^{2}-F/r^{2}-\rme^{2\phi}}},\\
\mathcal{P}(r)=\frac{\pi}{r^{2}}\int_{\rme^{\phi}}^{\infty}\rmd E\int_{0}^{r^{2}(E^{2}-\rme^{2\phi})}\rmd F\,\Phi(E,F)
\sqrt{E^{2}-F/r^{2}-\rme^{2\phi}},\\
\mathcal{P}_{T}(r)=\frac{\pi}{2r^{4}}\int_{\rme^{\phi}}^{\infty}\rmd E\int_{0}^{r^{2}(E^{2}-\rme^{2\phi})}\rmd F\frac{F\,\Phi(E,F)}
{\sqrt{E^{2}-F/r^{2}-\rme^{2\phi}}}.
\end{eqnarray}

A direct computation leads to the following
\begin{Lemma}
Assume $\phi,\rho,\mathcal{P},\mathcal{P}_{T}\in C^{1}([0,R))$, $R>0$. Then the following equation holds:
\begin{equation}
\mathcal{P}'(r)=-\rme^{2\phi}\phi'\rho(r)-\frac{2}{r}(\mathcal{P}(r)-\mathcal{P}_{T}(r)),\quad\forall r\in [0,R),
\end{equation}
which is know as ``Tolman-Oppenheimer-Volkov equation'' of a spherically
symmetric relativistic fluid ball (cf. \cite{W}, equation 3.2.19).
\end{Lemma}

\subsection{Existence of spherically symmetric steady states}
In this subsection we prove the existence of global solutions of NV$_{*}[r]$ (see theorem 2 below). We split the proof in four steps.

\vspace{0.5cm}
\noindent\textit{Step 1. The reduced problem}

\noindent Since for any solution of NV$_{*}[r]$ the particle density has the form (3.14), the problem of proving the existence of spherically
symmetric steady states can be reduced to the analysis of a single ordinary differential equation. For this purpose we substitute 
(3.14) into the right hand side of (3.6). After a transformation of variables we get
\begin{equation}
\frac{\rmd}{\rmd r}(\phi'r^{2})=H_{\Phi}(r,\rme^{\phi}),
\end{equation}
where
\begin{displaymath}
H_{\Phi}(r,u)=\pi u^{2}\int_{u}^{\infty}\rmd E\int_{0}^{r^{2}(E^{2}-u^{2})}\rmd F\frac{\Phi(E,F)}
{\sqrt{E^{2}-F/r^{2}-u^{2}}}.
\end{displaymath}

The recipe to find solutions of NV$_*[r]$ now consists in the following steps: fixing the functional $\Phi$, solving (3.19) and
defining $f=\Phi(E,F)$. Clearly, in order to carry all this out, some restrictions on $\Phi$ must be prescribed.

\vspace{0.5cm}
\noindent\textit{Step 2. Restrictions on $\Phi$}

\noindent The first restriction on $\Phi$ is dictated by the physical requirement of finite mass of the corresponding steady states. By
(3.15) we have
\begin{eqnarray}
M&=4\pi\int \rmd r  \rho(r)r^2\nonumber\\
&=4\pi^{2}\int_{0}^{\infty}\rmd r\int_{0}^{r^{2}(E^{2}-\rme^{2\phi})}
\rmd F\int_{\rme^{\phi}}^{\infty}\rmd E\frac{E^{2}\,\Phi(E,F)}{\sqrt{E^{2}-F/r^{2}-\rme^{2\phi}}}.
\end{eqnarray} 
The following lemma corresponds to theorem 2.1 of \cite{RR2}:
\begin{Lemma}
Let $\Phi:\mathbb{R}^{2}\to [0,\infty)$ be measurable and $\phi$ be a solution of (3.6) with $\phi(0)=\phi_{0}$.
Then $M<\infty$ implies $\phi_{\infty}=\lim_{r\to +\infty}\phi<\infty$, $\Phi(E,F)=0$ a.e. for $(E,F)\in 
(\rme^{\phi_{\infty}},+\infty)\times [0,+\infty)$ and the energy is finite:
\begin{equation}
\mathcal{E}=M+\int \rmd r\,r^{2}(\phi')^{2}< \infty.
\end{equation} 
\end{Lemma}
\noindent\textit{Proof: }Integrating (3.6) we get
\begin{displaymath}
\int_{0}^{\infty}\rmd r\,\frac{\rmd}{\rmd r}(\phi'r^{2})=\int_{0}^{\infty}\rmd r\,r^{2}\rme^{4\phi}\mu(r)\leq M.
\end{displaymath}
Since  $\phi'(0)=0$ for any $C^{1}$ solution of (3.6), this implies $M\geq \lim_{r\to +\infty}\phi'r^{2}$. 
Since the function $(\phi'r^{2})$ is increasing, then the limit in
the right hand member of the latter inequality exists and by the condition of finite mass has to be
\begin{equation}
\lim_{r\to +\infty}\phi'r^{2}<\infty.
\end{equation}
Thus in particular $\lim_{r\to +\infty}\phi'=0$ and since $\phi$ is increasing this implies $\phi_{\infty}<\infty$. Furthermore by
(3.20) we have
\begin{displaymath}
M\geq 4\pi^{2}\int_{\rme^{\phi_{\infty}}}^{\infty}\rmd E\, E^{2}\int_{0}^{\infty}\rmd F\,\Phi(E,F)\int_{\sqrt{F/(E^{2}-\rme^{2\phi_{\infty}})}}^{\infty}
\rmd r\frac{1}{\sqrt{E^{2}-\rme^{2\phi}}}.
\end{displaymath}
If $E>\rme^{\phi_{\infty}}$, the integral with respect to $r$ in the right hand side of the previous inequality diverges and so the mass 
would be infinite. Finally, (3.22) implies 
$\phi'(r)=o(r^{-1-\varepsilon})$, as $r\to +\infty,$ for $\varepsilon\in [0,1)$
and therefore $r^{2}(\phi')^{2}\sim r^{-2\varepsilon}=o(1/r)$, for $\varepsilon>1/2$ and this proves (3.21).
 $\  \Box$
\begin{Remark}
\textnormal{Given a solution $(f,\phi)$ with finite mass, the new solution $(\tilde{f}, \tilde{\phi})$ defined by
$\tilde{f}=\rme^{4\phi_\infty}f$, $\tilde{\phi}=\phi-\phi_\infty$ gives a steady state which is asymptotically flat, i.e. satisfying the boundary
condition $\lim_{r\to\infty}\tilde{\phi}(r)=0$. We notice that, unlike the Vlasov--Poisson and the Einstein--Vlasov system, in order to obtain
an asymptotically steady state from a solution with finite mass it is necessary to rescale the particle density (cf. the remark at the
end of section 2 in \cite{RR2}).}
\end{Remark}

\noindent\textit{Step 3. Global existence: set up}

\noindent In this step we give the basic set up to prove, in step 4, global existence and uniqueness for (3.19) 
when $\Phi$ has the form:
\begin{eqnarray}
\Phi(E,F)=\Psi(E)F^{k},\, E>0,\, F>0, \,k>-1/2,\nonumber\\
\Psi\in L^{\infty}((0,\infty))\textnormal{ is non-negative},\\
\exists E_{0}>0 \textnormal{ s.t. } \Psi(E)=0\textnormal{ for }E>E_{0},\nonumber
\end{eqnarray}
which was considered in \cite{Re,RR2}. We start by rewriting the relevant equations according to (3.23).

With $\Phi$ given by (3.23), equation (3.19) becomes
\begin{equation}
(r^{2}\phi')'=\pi c_{k,-\frac{1}{2}}r^{2k+2} \rme^{2\phi}h_{k+\frac{1}{2}}(\rme^{\phi}),
\end{equation}
where 
\begin{eqnarray}
h_{m}(u)=\int_{u}^{\infty}dE\,\Psi(E)(E^{2}-u^{2})^{m},\quad m>0,\\
c_{a,b}=\int_{0}^{1}s^{a}(1-s)^{b}ds=\frac{\Gamma(a+1)\Gamma(b+1)}{\Gamma(a+b+2)},\quad a>-1, b>-1,
\end{eqnarray}
where $\Gamma$ is the gamma function.
Moreover, the quantities (3.15)-(3.17) take the form
\begin{eqnarray}
\rho(r)=\pi r^{2k}c_{k,-\frac{1}{2}}g_{k+\frac{1}{2}}(\rme^{\phi}),\\
\mathcal{P}(r)=\pi r^{2k}c_{k,\frac{1}{2}}h_{k+\frac{3}{2}}(\rme^{\phi}),\\
\mathcal{P}_{T}(r)=\frac{\pi}{2} r^{2k}c_{k+1,-\frac{1}{2}}h_{k+\frac{3}{2}}(\rme^{\phi}),
\end{eqnarray}
where 
\begin{equation}
g_{m}(u)=\int_{u}^{\infty}dE\,\Psi(E)E^{2}(E^{2}-u^{2})^{m}=h_{m+1}+u^{2}h_{m}.
\end{equation}
We also notice that in virtue of the identity $x\Gamma(x)=\Gamma(x+1)$, we have
\begin{equation}
\mathcal{P}_{T}(r)=(k+1)\mathcal{P}(r)
\end{equation}
and therefore (3.12) and (3.18) become respectively
\begin{eqnarray}
\frac{1}{r^{2}}(\phi'r^{2})'=\rho(r)-(2k+3)\mathcal{P}(r),\\
\mathcal{P}'(r)=-\rme^{2\phi}\phi'\rho(r)+\frac{2k}{r}\mathcal{P}(r).
\end{eqnarray}
The following lemma states the regularity properties of the functions $g_{m}, h_{m}$ defined by (3.25), (3.30) and is proved in 
\cite{Re,RR2}.
\begin{Lemma}
If $m>-1$ then $g_{m},h_{m}\in C^{0}((0,+\infty))$. For $m>0$, $g_{m},h_{m}\in C^{1}((0,+\infty))$ and $\forall u>0$:
\begin{displaymath}
h'_{m}=-2m\,u\, h_{m-1},\quad g'_{m}=-2m\,u\,g_{m-1}.
\end{displaymath} 
\end{Lemma}

We notice that by lemma 3, the quantities $\rho,\mathcal{P}$ and $\mathcal{P}_{T}$ are $C^{1}$ in $[0,R)$ provided that $\phi$ is 
$C^{1}$ in $[0,R)$. The latter is then a sufficient condition for the validity of (3.33) in $[0,R)$.

\vspace{0.5cm}
\noindent\textit{Step 4. Global existence: proof}

\noindent Here is our main result about the existence of spherically symmetric stationary solutions of NV$_*$:
\begin{Theorem}
Let $\phi_{0}=\phi(0)\in\mathbb{R}$ and $\Phi$ of the form (3.23) be given. Then $NV_{*}[r]$ has a unique global solution  
$(\phi,f=\Phi(E,F))$. If the function $\Psi$ in (3.23) is $C^{1}$, then $f$
solves the stationary Vlasov equation (3.7)
in a classical sense.
\end{Theorem}
\noindent\textit{Proof: }We have to prove global existence for (3.24) or, equivalently, for (3.32). The former will be used to prove local
existence, the latter to extend this solution globally. Integrating once (3.24) we get
\begin{displaymath}
\phi'=\frac{\pi}{r^{2}}c_{k,-\frac{1}{2}}\int_{0}^{r}\rmd s\,s^{2k+2}\rme^{2\phi(s)}h_{k+\frac{1}{2}}(\rme^{\phi}).
\end{displaymath}  
For the function $u=\rme^{\phi}$ we then obtain the equation
\begin{displaymath}
u'=\frac{\pi}{r^{2}}c_{k,-\frac{1}{2}}u\int_{0}^{r}\rmd s\, s^{2k+2}u^{2}h_{k+\frac{1}{2}}(u).
\end{displaymath}
Integrating the previous equation with the initial condition $u_{0}=u(0)=\rme^{\phi_{0}}>0$  we get
\begin{eqnarray}
u(r)&=u_{0}+\pi c_{k,-\frac{1}{2}}\int_{0}^{r}\rmd s\frac{u(s)}{s^{2}}\int_{0}^{s}\rmd\tau\,\tau^{2k+2}u^{2}(\tau)h_{k+\frac{1}{2}}(u(\tau))\nonumber\\
&:=(Tu)(r)\nonumber.
\end{eqnarray}
Now consider the set $M_{\delta}=\{u:[0,\delta]\to [0,+\infty)\,\textnormal{ continuous}:u_{0}\leq u(r)\leq u_{0}+1\}$.
By a simple calculation one can show that, for $\delta$ small enough, the operator $T$ maps $M_{\delta}$ onto itself and that 
$T:M_{\delta}\to M_{\delta}$ is a 
contraction with respect to the $L^{\infty}$ norm. This gives local existence for (3.24). 
Standard techniques permit now to extend $\phi$ to a maximal $C^{1}$ solution on an interval $[0,R)$, $R>0$. If $R<\infty$ then
\begin{displaymath}
\lim_{r\to R^{-}}\phi(r)=\infty\quad\textnormal{and/or}\quad\lim_{r\to R^{-}}\phi'(r)=\infty.
\end{displaymath}
Assume 
\begin{equation}
\lim_{r\to R^{-}}\phi(r)=\alpha<\infty. 
\end{equation}
Then
\begin{eqnarray}
\int_{0}^{R}\rmd r\, \frac{\rmd}{\rmd r}(\phi'r^{2})&=\pi c_{k,-\frac{1}{2}}\int_{0}^{R}\rmd r\,
r^{2k+2}\rme^{2\phi(r)}h_{k+\frac{1}{2}}(\rme^{\phi})\nonumber\\
&\leq \pi c_{k,-\frac{1}{2}}R^{2k+3}\rme^{2\alpha}h_{k+\frac{1}{2}}(\rme^{\phi_{0}}),\nonumber
\end{eqnarray}
where we used the fact that $h_{m}$ is decreasing and $\phi$ is increasing.
Hence,
\begin{displaymath}
\lim_{r\to R^{-}}\phi'\leq CR^{2k+1}<\infty,
\end{displaymath}
where $C$ is a positive constant.
Thus it is sufficient to prove (3.34). For this purpose, we notice that by (3.32) we have $(\phi'r^{2})'\leq r^{2}\rho(r)$.
Since $\phi'>0$, integrating twice this inequality we get
\begin{eqnarray}
\phi(r)&\leq\phi_{0}+\int_{0}^{r}\frac{\rmd s}{s^{2}}\int_{0}^{s}\rmd\tau\,\tau^{2}\rho(\tau)\nonumber\\
&\leq\phi_{0}+\pi c_{k,-\frac{1}{2}}\int_{0}^{r}\frac{\rmd s}{s^{2}}\int_{0}^{s}\rmd\tau\,\tau^{2k+2}g_{k+\frac{1}{2}}(\rme^{\phi})\nonumber\\
&\leq \phi_{0}+Cr^{2k+2},\nonumber
\end{eqnarray}
by which (3.34), and then global existence, follows. $\  \Box$

\begin{Remark}
\textnormal{Since $g_{m}(u), h_{m}(u)=0$, for $u\geq E_{0}$ and $\phi$ is increasing, then for
$\phi_{0}\geq\log E_{0}$, 
the unique solution of theorem 2 is given by $\phi(r)=\phi_{0}$, $\forall r>0$ and $f=0$ a.e.}
\end{Remark}

\subsection{Steady states with finite radius}
In this section we face the problem whether, among the solutions of theorem 2, there are steady states with compact support, i.e. such
that
\begin{equation}
\exists R\in (0,+\infty):\,\rho(r)=0,\quad\textnormal{for } r>R.
\end{equation}
We start by pointing out that a steady state with finite radius has also finite mass (and so finite total energy, cf. lemma 2). For
\begin{eqnarray}
M&=\int_{0}^{R}\rmd r\, r^{2}\rho(r)=\pi c_{k,-\frac{1}{2}}\int_{0}^{R}\rmd r\, r^{2k+2}g_{k+\frac{1}{2}}(\rme^{\phi})\nonumber\\
&\leq \pi c_{k,-\frac{1}{2}}g_{k+\frac{1}{2}}(\rme^{\phi_{0}})\frac{R^{2k+3}}{2k+3}<\infty.\nonumber
\end{eqnarray}

Since by (3.27), $\rho(r)=0$ if and only if $\rme^{\phi(r)}\geq E_{0}$, then the problem of proving (3.35) is equivalent to show the existence 
of $R\in (0,+\infty)$ such that $\rme^{\phi(r)}=E_{0}$. For this purpose we will need later the following simple lemma:
\begin{Lemma}
Let $\phi\in C^1$ and $[0,R)$ be the maximal interval such that $\rme^{\phi(r)}<E_{0}$. Then 
\begin{displaymath}
\lim_{r\to R^{-}}\rme^{\phi(r)}=E_{0}.
\end{displaymath}  
\end{Lemma}
\noindent\textit{Proof: }If $R<\infty$ this is obvious. For $R=+\infty$, let $\rme^{\phi_{\infty}}=\lim_{r\to\infty}\rme^{\phi(r)}\leq E_{0}$
and assume $\rme^{\phi_{\infty}}<E_{0}$. Then by (3.24), $(r^{2}\phi')'\geq C\, r^{2k+2}$, which implies, upon integration, $\phi'\to\infty$,
and so $\phi\to\infty$, for $r\to +\infty$, a contradiction. $\   \Box$

We will prove the finite radius property for steady states of the form (3.23), where $\Psi$ satisfies the additional condition 
\begin{equation}
\Psi(E)=c(E_{0}-E)^{\mu}_{+}+O((E_{0}-E)_{+}^{\mu+\delta}),\quad\textnormal{as } E\to E_{0}^{-},
\end{equation}
where $(\cdot)_{+}$ denotes the positive part, $c,\delta>0$ and 
\begin{eqnarray}
\mu >-1,\quad k>-\frac{1}{2},\\
\frac{2\mu+1}{\mu+k+5/2}<E_{0}^{2}<\frac{2\mu+2k+3}{\mu+k+5/2}.
\end{eqnarray}

This class was considered in \cite{RR2}, except for the condition (3.38) which is different in \cite{RR2}. 
We will prove our result following the argument of \cite{RR2}. The result is the 
following:
\begin{Theorem}
Let $(f,\phi)$ be a solution of NV$_*$[$r$] of the form stated above. Then this steady state has finite radius.
\end{Theorem}
\noindent\textit{Proof: }We consider only the non trivial case $\phi_{0}<\log E_{0}$ (cf. remark 3). Define
\begin{eqnarray*}
q(r)=\rho(r)-(2k+3)\mathcal{P}(r),\\
m(r)=\int_{0}^{r} \rmd s\,s^{2}q(s),\quad\eta(r)=\log E_{0}-\phi(r).
\end{eqnarray*}
We notice that $\eta(r)>0$ for $r<R$ and $\eta\to 0$ for $r\to R^{-}$, where $R$ is defined as in lemma 4.
We also introduce the functions:
\begin{displaymath}
x(r)=\frac{m(r)}{r\eta(r)},\quad\quad y(r)=r^{2}\frac{q^{2}(r)}{\mathcal{P}(r)}.
\end{displaymath}
Since $r\eta'=-\eta x$, then 
\begin{equation}
rx'=-x+x^{2}+\alpha (r)y,
\end{equation}
where 
\begin{equation}
\alpha(r)=\frac{\mathcal{P}}{\eta q}=(2k+3)^{-1}\frac{h_{k+\frac{3}{2}}(\rme^{\phi})}{\eta \rme^{2\phi}h_{k+\frac{1}{2}}(\rme^{\phi})}.
\end{equation}
Using (3.27), (3.33) and lemma 3 we also  get
\begin{equation}
ry'=(2+2k)y-\beta(r)xy,
\end{equation}
where 
\begin{eqnarray}
\beta(r)=&-(2k+3)\eta \rme^{2\phi}\frac{g_{k+\frac{1}{2}}(\rme^{\phi})}{h_{k+\frac{3}{2}}(\rme^{\phi})}
-2(2k+3)\eta\frac{g_{k+\frac{1}{2}}(\rme^{\phi})}{h_{k+\frac{1}{2}}(\rme^{\phi})}\nonumber\\
&+2(2k+1)\eta\frac{g_{k-\frac{1}{2}}(\rme^{\phi})}{h_{k+\frac{1}{2}}(\rme^{\phi})}.
\end{eqnarray}
We reduced the field equation to the form (3.39), (3.41), which is the same stated in lemma 3.2 of \cite{RR2} with $\gamma_{1}=\gamma_{2}=0$.
Thus if we prove that
\begin{itemize}
\item[A)] $\alpha_{0}=\inf_{r\in (0,R)}\alpha (r)>0$
\item[B)] $\beta_{0}=\lim_{r\to R^{-}}\beta(r)\in (0,2+2k)$,
\end{itemize}
we conclude that $R<\infty$.

By means of (3.36) we have
\begin{displaymath}
\Psi(E)=c_{*}E(E_{0}^{2}-E^{2})^{\mu}_{+}+E\, O((E_{0}^{2}-E^{2})_{+}^{\mu+\delta}),\textnormal{ as } E\to E_{0}^{-}.
\end{displaymath}
We may assume $c_{*}=1$, because this constant cancels in $\alpha$ and $\beta$. Using the relation
\begin{eqnarray}
\int_{u}^{E_{0}} E (E_{0}^{2}-E^{2})^{a}(E^{2}-u^{2})^{b}\rmd E=\frac{1}{2}c_{a,b}(E_{0}^{2}-u^{2})^{a+b+1},\nonumber
\end{eqnarray}
which is valid for $u\leq E_{0},\, a>-1,\, b>-1$ we get
\begin{displaymath}
h_{m}(\rme^{\phi})=\frac{1}{2}c_{\mu,m}\varepsilon^{\mu+m+1}+O(\varepsilon^{\mu+m+\delta+1}),\quad \varepsilon=E_{0}^{2}-\rme^{2\phi}\to 0^{+}.
\end{displaymath}
Moreover, since $\eta=\log(1+\rme^{-\phi}\varepsilon/(E_{0}+\rme^{\phi}))$, then
\begin{displaymath}
\eta=\frac{1}{2E_{0}^{2}}\varepsilon +O(\varepsilon^{2}),\quad \varepsilon\to 0^{+}.
\end{displaymath}
Thus, since $\varepsilon\to 0^{+}$ for $r\to R^{-}$ (see lemma 4), we have
\begin{eqnarray}
\lim_{r\to R^{-}}\alpha (r)&=(2k+3)^{-1}\frac{c_{\mu,k+\frac{3}{2}}}{c_{\mu,k+\frac{1}{2}}}\nonumber\\
&=(\mu+k+\frac{5}{2})^{-1}>0.\nonumber
\end{eqnarray}
Since $\alpha>0$ on $(0,R)$ and $\lim_{r\to 0^{+}}\alpha(r)>0$ (because of our assumption $\phi_{0}<\log E_{0}$), then A) is proved.
To prove B) we notice that
\begin{eqnarray}
g_{m}(\rme^{\phi})&=h_{m+1}+\rme^{2\phi}h_{m}\sim\rme^{2\phi}h_{m}\nonumber\\
&=\frac{E_{0}^{2}}{2}c_{\mu,m}\varepsilon^{\mu+m+1}+O(\varepsilon^{\mu+m+\delta+1}),\  \ \varepsilon\to 0^{+}.\nonumber
\end{eqnarray}
This immediately implies that the second term of (3.42) tends to zero as $\varepsilon\to 0^{+}$ and therefore we get
\begin{eqnarray}
\lim_{r\to
R^{-}}\beta(r)&=-\frac{E_{0}^{2}}{2}\frac{c_{\mu,k+\frac{1}{2}}}{c_{\mu,k+\frac{3}{2}}}+\frac{2k+1}{2k+3}\frac{c_{\mu,k-\frac{1}{2}}}
{c_{\mu,k+\frac{1}{2}}}\nonumber\\
&=-E_{0}^{2}(\mu+k+\frac{5}{2})+2\mu+2k+3,\nonumber
\end{eqnarray}
which lies in the required interval $(0,2k+2)$ by our assumptions on $E_{0},\mu,k$. $\   \Box$

\ack{
I thank Alan Rendall for many helpful discussions and Gerhard Rein for useful comments on the manuscript. This research was partially
supported by the European network HYKE (contract HPRN-CT-2002-00282).}


\end{document}